%% file: hotnets15.tex
\documentclass{hotnets15}

\usepackage{times}  
\usepackage{epsfig}
\usepackage[TABBOTCAP]{subfigure}
\usepackage{tabularx}
\usepackage{graphicx} 
\usepackage{color}
\usepackage{xspace}
\usepackage{thumbpdf}
\usepackage{listings}
\usepackage{verbatim}
\usepackage{hyperref}
\usepackage{booktabs}
\usepackage{colortbl}

\hypersetup{pdfstartview=FitH,pdfpagelayout=SinglePage}

\begin{document}

\conferenceinfo{HotNets 2015} {}
\CopyrightYear{2015}
\crdata{X}
\date{}


\title{Quality of Consumption:\\The Friendlier Side of Quality of Service}

\author
{
Murad Kablan$^\dagger$
\and Hani Jamjoom$^\ddagger$ 
\and Eric Keller$^\dagger$
\end{tabular}\newline\begin{tabular}{c c c}
        \affaddr{$^\dagger$University of Colorado} & ~~~~~ & \affaddr{$^\ddagger$IBM Watson Research Center}\\
       \affaddr{Boulder, CO, USA} & ~~~~~ & \affaddr{Yorktown Heights, NY, USA}\\
}

\newcommand{\paragraphb}[1]{\vspace{0.03in}\noindent{\bf #1} }
\newcommand{\paragraphe}[1]{\vspace{0.03in}\noindent{\em #1} }
\newcommand{\paragraphbe}[1]{\vspace{0.03in}\noindent{\bf \em #1} }
\newcommand{\eric}{\textcolor{red}}
\newcommand{\murad}{\textcolor{blue}}

\newcommand{\eg}{{\em e.g., }}
\newcommand{\ie}{{\em i.e., }}
\newcommand{\fixme}[1]{\textit{\textcolor{red}{[#1]}}}
\newcommand{\findcite}[1]{\textcolor{magenta}{$^\text{[}$\footnote{\textcolor{red}{Find citation: #1}}$^\text{]}$}}
\newcommand{\nip}[1]{\vspace{1ex}\noindent\textbf{#1}}

\maketitle


\input{abstract}
\input{intro}

\input{defining-qox}

\input{sharing}

\input{provider-vouching}

\input{personalization}
\input{impl}
\input{conclusion}

\bibliographystyle{abbrv} 
\begin{small}
\bibliography{hotnets15,refs,ibm}
\end{small}
\label{last-page}

\end{document}

%% file: abstract.tex
\begin{abstract}
Cloud services today are increasingly built using functionality from
other running services. In
this paper, we question whether legacy Quality of Services (QoS)
metrics and enforcement techniques are sufficient as they are producer
centric. We argue that, similar to customer rating systems found in
banking systems and many sharing economy apps (e.g., Uber and Airbnb),
{\em Quality of Consumption (QoC)} should be introduced to capture
different metrics about service consumers. We show how the combination of QoS and QoC, dubbed QoX, can be used by consumers and providers to improve the security and management of their infrastructure. In addition, we demonstrate how sharing information among other consumers and providers increase the value of QoX. To address the main challenge with sharing information, namely sybil attacks and mis-information, we describe how we can leverage cloud providers as vouching authorities to ensure the integrity of information.  We present initial results in
prototyping the appropriate abstractions and interfaces in a
cloud environment, focusing on the design impact on both service providers and consumers.

\end{abstract}

%% file: intro.tex
\section{Introduction}

Building and deploying any distributed ``app'' today is radically
different from a decade ago. Where traditional applications of the
past required dedicated infrastructure and middleware stacks, today's
apps not only run on shared---cloud---infrastructure, they rely on
many services, residing within and outside of underlying cloud. An
app, for example, can use Facebook for authentication, Box for
storage, Twilio for messaging, Square for payments, Google AdSense for
advertising, etc. This trend of deploying and consuming services
(often referred to as the {\em mesh economy}) can be seen by the rapid
growth of cloud platforms which integrate services (and not just
compute) like Amazon Web Services~\cite{aws}, Microsoft
Azure~\cite{windowsazure}, Heroku~\cite{heroku}, IBM
BlueMix~\cite{bluemix}, and CloudFoundry~\cite{cloudfoundry}, to name
a few. More importantly, these emerging platforms further encourage
the developed apps to expose their core capability as services to be
consumed by other services.  The result is a growing ecosystem of
interdependent services that blur the traditional boundaries between
service producers and consumers.

In this paper, we revisit quality of service (QoS) abstractions and
enforcement techniques in such environments. Specifically, we observe
that while QoS is the subject of much research, it has been primarily
provider centric. For example, QoS for video streaming looks at the
quality of provider's video and how it is impacted by bandwidth, latency,
jitter, etc. This unidirectional view has dominated the design of past
QoS abstractions and enforcement techniques. Even today, most cloud
services continue to design their APIs around static entitlement
buckets (e.g., free clients are allowed 100 calls per minutes and
registered users are allowed 1000 calls per minute). If viewed as a
directed graph,\footnote{In such graph, service producers and
  consumers are represented as nodes, and consumption from a producer
  to a consumer is represented as a directed edge} current QoS models
focus on either a single interaction between a producer and consumer
(e.g., rate limits) or producer and many consumers (e.g., DoS
protection). In this paper, we study how interactions among all
producers and consumers can be leveraged to the general benefit of
everyone.

Thinking of the problem space from an ecosystem perspective creates
new opportunities for improving both service production and
consumption. In this new environment, better service production is encouraged. Similarly, better service consumption is
rewarded. Finally, poorly implemented services and malicious clients
are isolated. Achieving this vision requires questioning the
unidirectionality of QoS designs. In particular, we introduce {\em
  Quality of Consumption (QoC)} to refer to metrics that can be
captured to define attributes of how a consumer is using a service.
The idea of QoC is not new in the real world. Lending (e.g, mortgage
and credit cards) are dependent on the ratings of the
customers. Customers (credit consumers) who demonstrate consistent
repayment of loans have improved FIMCO scores; this, in turn results
in higher future credit limits.

In most real systems, both QoS and QoC are needed. Recent startups in
the {\em sharing economy} space, like Uber and Airbib, demonstrate how
quality of the providers and consumers can be leveraged to enable
trust between all parties. We use the term QoX to capture systems that
integrate both QoS and QoC. Conceptually, there are then three import
components to QoX environments: (C1) providers' QoS are monitored and
rated, (C2) consumers' QoC are similarly tracked, and (C3) this
information is shared among providers and consumers and is used in
provider selection and service differentiation.

In this paper, we show how a similar setup can be achieved in cloud
environments. Specifically, we look at how to extend existing QoS
frameworks to support QoC. This is to support C1 and C2
above. Furthermore, we also look at how to use cloud providers as
vouching authorities to achieve C3, even in the presence of sybils and
liers.
%
%
We present initial thoughts on the appropriate abstractions and
interfaces to address them on a cloud based framework that manages and
define the quality of interaction and service from both consumer and
provider's perspectives. We explore the motivations, challenges, and
potentials to introduce such a framework in the cloud environment.

%% file: defining-qox.tex
\section{Defining Quality}


In this section, we introduce the term Quality of Consumption (QoC) as
a counterpart to the Quality of Service (QoS) metrics. Both can be
readily measured (latency, bandwidth, etc.) and can be specified as
part of SLAs. We define QoS and QoC. We also describe the needed
components to enable QoX in cloud environments.


\nip{Quality of Service (QoS).} QoS can be defined as a measurable level of
service delivered to its users'
satisfaction.  QoS of cloud services can be characterized across
multiple dimensions, each having a set of metrics
(Table~\ref{tab:qos-metrics}).  For example, the quality of a database
service, such as ClearDB\cite{clearDB}, can be measured by its dependability
(availability) and performance (response time for SQL request). While
the quality of an ad service such as Adobe Ad\cite{adobe-ad} can be measured by its
rate.

\begin{table}
\centering
\small
\begin{tabular}{|  p{1.5cm} |  p{5cm} |}
  \hline                       
  \rowcolor[gray]{0.9} {\bf Dimension} & {\bf Metrics} \\
  \hline
  Performance & Throughput, Packet loss probability, response time, jitter\\
  \hline
  Dependability & Reliability (\eg maximum number of crashes or interruption), availability (\eg maximum number the service will be unavailable)\\
  \hline
  Cost & Prices and rates\\
  \hline
\end{tabular}
\caption{QoS metrics of Cloud services}
\label{tab:qos-metrics}
\end{table}

\nip{Quality of Consumption (QoC).} QoC captures how well users are
consuming a service. It can be used by service providers to assist in
admission control decisions or when providing service differentiation.
QoC is a way to recognize that service consumers are not equal.  To
some extent, QoC monitoring already exists (e.g., intrusion detection
and prevention systems).  These are point solutions.  The problem is
that there is no abstraction or framework for cloud service providers
to collect various QoC metrics and then translate them to suitable
actions.  Similar to QoS, QoC can be characterized across multiple
dimensions, each having a set of metrics
(Table~\ref{tab:qoc-metrics}).

\begin{figure*}
    \centering
    \includegraphics[width=.85\textwidth]{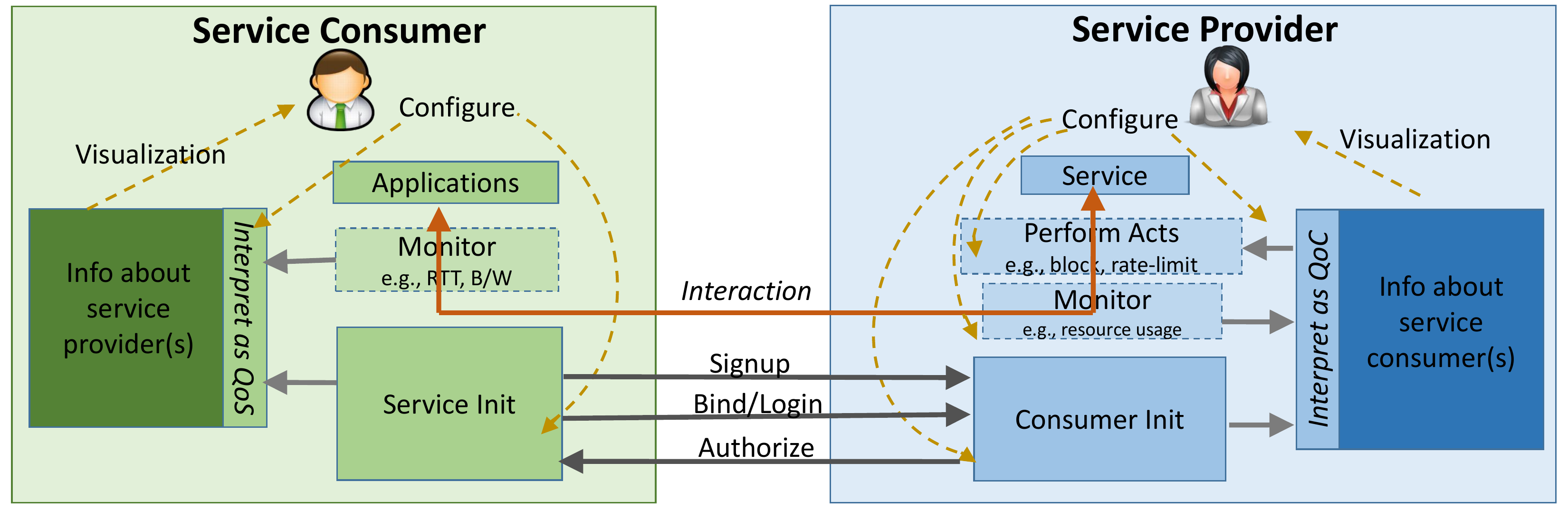}
    \caption{\textsf{Interpretation of measured information as quality of service or quality of consumption.}}
    \label{fig:qox-defined}
\end{figure*}

\begin{table}
\centering
\small
\begin{tabular}{|  p{1.7cm} |  p{5cm} |}
  \hline                       
  \rowcolor[gray]{0.9} {\bf Dimension} & {\bf Metrics} \\
  \hline
  Customer purchase power &  Length existing as a user, frequency of orders, amount of purchases \\
  \hline
  Customer's code efficiency  & Version of software customer is running, malformed requests (\eg Web server error logs)\\
  \hline
  Customer threat & IDS alerts. Service crash reports\\
  \hline
\end{tabular}
\caption{QoC metrics of Cloud services}
\label{tab:qoc-metrics}
\end{table}


\nip{QoX = QoS + QoC.} We use the abbreviation QoX to capture the
combination of QoS and QoC. As illustrated in Figure~\ref{fig:qox-defined},
through measurement systems and other system logs, consumers and providers capture 
quality metrics about each other (\eg via an IDS or other resource monitors), captured in the box labeled 
information about service provider(s) or consumer(s) -- indicating it is available, not necessarily stored. 
Collectively, this information is interpreted as QoX, and can be used, either directly
or indirectly, in managing the infrastructures operation, both at run-time and during initialization.
%
%
We discuss the details of
implementation and integration in Section~\ref{sec:impl}.

%% file: sharing.tex
\section{Democratization and Sharing of QoX}
\label{sec:sharing}

In sharing information, consumers and providers can gain the benefit
of others' experiences. 
This can be useful, for example, when choosing a service provider, or
knowing that certain consumers are likely to pose security threats.
Even more, this information can be used for self-feedback.  In
traditional infrastructures, the administrator has visibility of
what is happening inside of the infrastructure through a variety of
monitoring tools.  The administrator, however, has limited visibility
into how the infrastructure is viewed externally.  
Sentiment analysis is widely used in corporations (\eg
monitor Twitter feeds to observe whether there is any positive or
negative chatter affecting its brand~\cite{henshen2012, o2010tweets}).
With an information exchange system, a provider (or consumer) can
monitor its sentiment as perceived by its consumers (or providers),
and trigger a root cause analysis if there are any negative issues.

\begin{figure*}
	\centering
	\includegraphics[width=.85\textwidth]{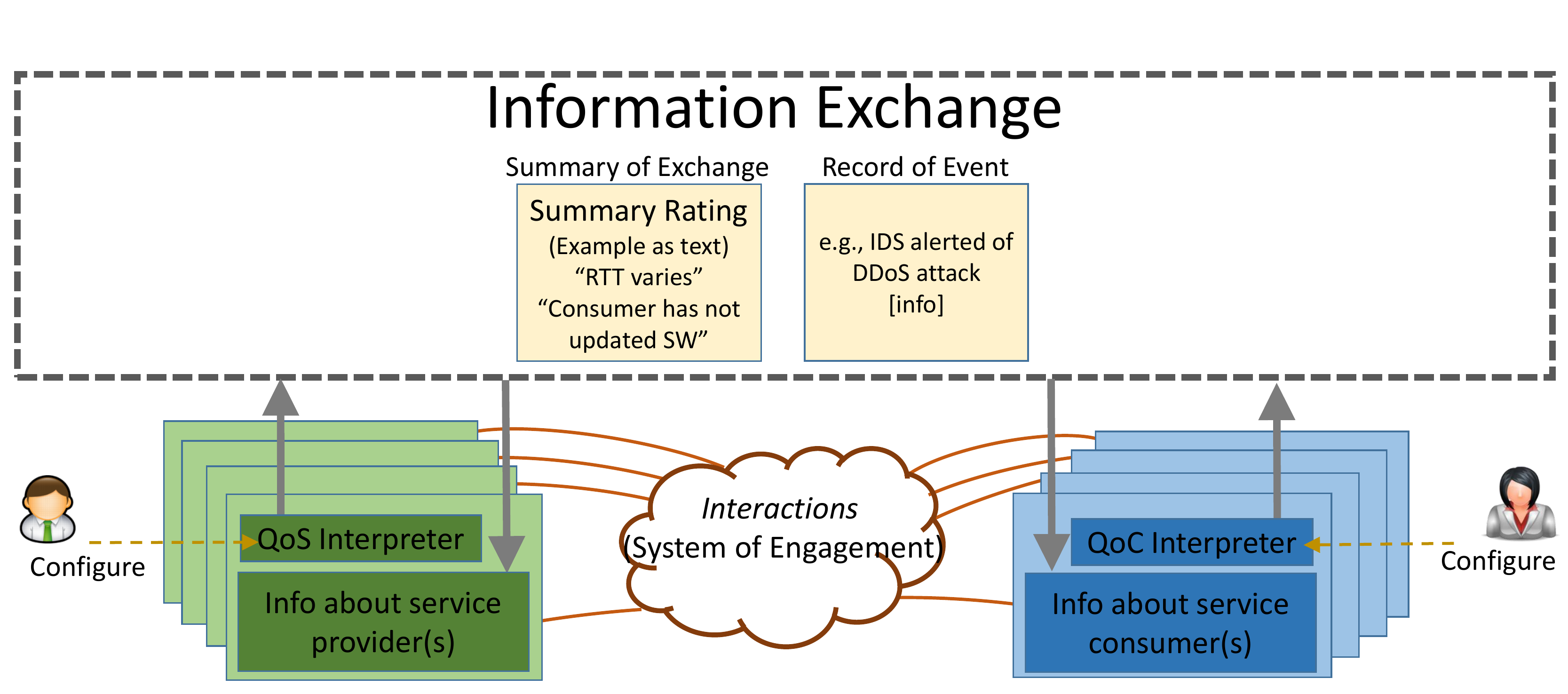}
	\caption{Illustration of information exchange}
	\label{fig:qoxsharing}
\end{figure*}

Illustrated in Figure~\ref{fig:qoxsharing} is an ecosystem of consumers
and providers, all interacting with one another (forming a system of
engagement~\cite{systemofexchange}), and exchanging the information.
The information about a service (or consumer), previously illustrated
in Figure~\ref{fig:qox-defined}, is shared with a logical service
labeled information exchange.  In the remainder
of the section, we discuss the two main types of information. The
first summarizes information about the interaction as a whole; the
second is a record of a specific interaction.  We address the challenges
of dealing with lying and sybils in Section~\ref{sec:vouching}.

\subsection{Summary of Engagement}

The challenge in simply exchanging the information about a provider
(or consumer) is there is no clear, standardized way to compare
quantifiable metrics across providers and consumers.  
Even simple and as well-defined metrics such as latency can be
subjective (\eg due to network proximity). 
%
Instead, we are inspired by review systems found in web sites such as
Yelp (for restaurants) and Amazon (for products).  In this case,
consumers and providers share a scalar, subjective rating of quality
coupled with a (machine generated) text based review.
%


\subsubsection{Scalar, Subjective Rating}

The summarization rating is scalar (as opposed to just
good or bad) as there are many factors that go into overall quality, and
subjective to account for the wide variety of metrics and needs of
various providers or consumers.  This rating represents the current
view, with the weighting of history versus recent experiences left to
the rater.
%
%
A key challenge is determining the rating.  As illustrated in
Figure~\ref{fig:qoxsharing}, we envision that each consumer and
provider will have software which interprets the information about a
consumer or provider to ultimately determine the rating .  The
administrator configures how this interpreter behaves.  This is a long
term challenge---creating a language to enable administrators on
either side to integrate with the measurement systems, being able to
specify expectations, and how each component impacts the overall
rating.  For initial exploration, we can provide thresholds for
specific metrics, or simply let the human administrators provide a
rating.

In
Section~\ref{sec:impl}, we will briefly discuss how consumers and
providers can use QoX information to help automatically manage their
infrastructure (\eg load balancing or prioritization based on QoC).
Having the shared information can strengthen the confidence in those
actions, but also opens some additional avenues for the specific case
of a given consumer and provider having not had interaction before.
On the consumer side, the shared information can support the consumer
in making a decision about which service provider to use.  On the
service provider side, the shared information can support the provider
in offering incentives to use a service or introducing restrictions in
use.

\subsubsection{Text review to Help Interpret the Rating}

A coarse-grained rating has the benefit that it can be used in any
context.  The downside, however, is that it hides potentially useful
information (\eg information about why a particular reviewer gave the
rating they did).  This introduces a challenge in interpreting the
rating.  Using a restaurant example, consider a restaurant that has
several 2-star ratings and several 5-star ratings.  The 2-star ratings
ultimately reflect that those diners really value authenticity, and
this restaurant was not authentic enough, whereas the 5-star ratings
ultimately reflect that while it might not be really authentic, the
food is tasty and the restaurant is really clean and has friendly
staff.  This distinction cannot be captured with a coarse rating.

One approach to getting around this is to use sub-categories, though
this has notable downside of determining the sub-categories requires
forecasting every single criteria that might be used---an impossible
task.  In practice, a small set of sub-categories is useful (\eg Home
Depot has quality and value in addition to overall rating, to help
separate all metrics related to the product and metrics related to the
cost), but having too many will inevitably require standardization.

Instead, we propose including machine-generated text based reviews to go along with the
rating.  This will allow each reviewer the means to specify why they
gave the rating, and each user of a review to find whether the rating
reflects its needs.  Clearly, text-based reviews are helpful for human
adminstrators if they want to view the ratings, but we believe text
reviews can also help guide automated systems as well.

Through systems (such as Elastic Search) which extract structure from
unstructured data and provide analysis, we believe that it will be
possible to extract the commonalities among reviews -- that is,
automatically creating sub-categories that are relevant for that
particular provider or consumer.  Amazon does this for product reviews
as they highlight common comments.

Creating these machine-generated text reviews is not a significant
challenge as they can be built out of specific log text, and based on
the administrators configuration of the QoX interpreter.  Configuring
the infrastructure to search for specific information will require
some administrative effort initially, to understand, generally, what
the commonalities are and then building that into the configuration
for how to use the ratings as they are adjusted overtime.

\subsubsection{Personalization without Connections}
In other information exchange systems, there have been proposals to
leverage social structure to provide more relevant information.  For
example, in a system which rates restaurants, any reviews from friends
would be more trusted/valued and more useful/relevant.  In short,
personalization arranges the presentation of, filters, and interprets
the information to be most relevant to the requester of the
information.

We believe the same is true for cloud services, except, of course,
there is no notion of friends among service providers or consumers.
Previous attempts~\cite{Frazier2011} have extracted such a `social'
structure by way of interactions.  This, however, requires a fairly
connected graph.  Whereas, the consumer-provider graph is likely
to be mostly a bi-partite graph.\footnote{This is also a challenge we
  highlighted in using existing sybil attack prevention approaches.}

Instead, we can perform personalization by leveraging techniques used
by recommendation engines.  In recommendation systems, the goal is to
predict a choice (\eg what movie to watch) based on others with like
characteristics (\eg who have watched similar movies).  That is, they
find connections between unconnected users using machine learning
techniques (\eg PredictionIO~\cite{predictionio}).

Our goal is not to make a recommendation, but to highlight reviews
that are particularly relevant.  The challenge is identifying relevant
features.  This may include examining consumers who use a similar set
of services, or who uses a similar set of API calls for a given
service.  This is an area for future research.

\subsection{Exchanging Specific Records}


Ultimately, exchanging summary of exchange information will capture a scalar rating of different metrics about
consumers and providers. While ratings, even detailed ratings, have value in making coarse
decisions, some information that an individual consumer or provider
records would help others if shared.  This is specifically evident in
securing an infrastructure.  At a high level, consider the
possibilities if each service provider share their intrusion detection
system (IDS) logs and alerts.  Now, all providers could get the
benefit of a `global IDS', allowing them to protect themselves even
before seeing an attack and even if they didn't have their own ability
to detect a given attack.

As a specific example, in 2014, hackers exploited a bug in the
Amazon EC2 API to gain access to other tenants' accounts and then flood other servers with UDP 
packets to cause a denial of service~\cite{ec2dos1, ec2dos2}. 
Providers sharing alerts of the attack, can greatly reduce the damage to services 
by warning others of potential threats, allowing them to, for example, block that particular 
tenant.  Note that in this example, while it may not prevent the attack from successfully denying service for
the first tenant attacked, it generally helped others (making it valuable), and there are a larger set of 
situations where everybody benefits -- specifically in detecting, and comparing, reconnaissance efforts
of potential attackers.

%% file: provider-vouching.tex
\section{Cloud Provider as Vouching Authority}
\label{sec:vouching}

A key challenge arises when dealing with any sharing is ensuring the
validity of information.  As we move toward an entirely cloud based
infrastructure---both IaaS and PaaS---we believe that an opportunity to
overcome these challenges becomes possible, where the cloud provider serves as a
vouching authority.  We elaborate on two ways by which validity of
information can be compromised, and the role of the cloud provider in
each.

\subsection{Fake Identities} 

In Section~\ref{sec:sharing}, we proposed that each consumer or
provider can rate the other party for which they have had an
interaction with.  In an ideal world, quality is computed by equally
weighting everyone's opinion (e.g., via average rating).  In reality,
reviewers can create fake identities, also know as the \emph{sybil
  attack}~\cite{Douceur2002}.  For example, if a service provider were
able to create a substantial number of fake entities that then rate
its service as high quality, that service provider's rating will be
unnaturally high. The inverse is possible, where a competitor that
wants to negatively influence another provider's rating.  There are a
number of prevention techniques~\cite{mislove2008ostra,sybilguard} that have
been designed for decentralized (peer-to-peer) systems. These will not
work here as each relies on a trust graph (\eg a social network graph)
in determining the likelihood that an entity is real and limiting the
influence the collection of sybils can have. Such graph does not exist
here.

Fortunately, these are not necessary for cloud based interactions as
we have a central authority: the cloud provider.  The cloud provider
can vouch for the identity is real and unique.  It is much more
challenging to create fake accounts where identity verification is
required, such as requiring a credit card (as Amazon does).  While it
may be possible for a small-scale attack by creating a few accounts
(with a few credit cards), overall this prevention mechanism is
sufficient if the cloud provider can vouch~\cite{sybilguard}



\subsection{Providing mis-information}

While having the cloud provider vouch for an identity prevents sybil
attacks, real parties can provide false information.  Here, we propose
that the cloud provider can, to some degree, vouch for the validity of
information based on the visibility the cloud provider has.

\nip{Did the reviewer actually interact with the reviewee?}
Generally, we can rely on the crowd to deal with any lying, as a
single rating will be noise in the overall rating.  For example, if
everyone is giving a service or consumer 5 stars except for its main
rival, then not only will that rating have little ability to influence
the overall score, but it might stand out and hurt the party giving
that low rating.  However, it is still desirable to restrict the
ability to rate parties that one has not interacted with, to prevent
collusion or compromises in some accounts.  An IaaS provider has
visibility into the network infrastructure, so can, for example, see
whether the tenants exchanged at least a certain number of packets.
An PaaS provider, such as running CloudFoundry~\cite{cloudfoundry},
brokers the interconnection between service provider and consumer, so
has the ability to indicate if a connection was actually made.

\nip{Can the information be trusted enough to act on immediately?}
Some information would cause immediate, automated action and as such
lying can have a negative impact.  Consider one service provider
detecting the start of some attack (\eg a DoS attack), sharing this
information with another provider will help them, say block that user.
If they are able to lie about it, they can cause another service
provider to block a consumer unnecessarily.  So, the receiver of the
information needs to ensure the information is accurate.

Again, the visibility the cloud provider has can be leveraged to
validate certain information.  Of course, this is a greater challenge
to deal with than simply determining if two parties interacted.  The
challenge lies in the variety of information that can be shared.  Each
will have different characteristics which will serve as evidence (\eg
if one tenant wishes to share that another tenant performed a port
scan, then the cloud provider needs enough evidence to verify that
occurred).  For this, we envision the tenant pre-specifying
\emph{evidence patterns}, which will specify the evidence that the
tenant would like the cloud provider to collect. We envision
measurable information such as bursts of traffic, crashes, specific IP
address did indeed send something to another IP address, and not
performing deep packet inspection, but this is an area for future
investigation.

%% file: impl.tex
\section{Initial Prototype}
\label{sec:impl}

As an initial proof-of-concept prototype, we focus on demonstrating the feasibility
of integration with existing systems and applications.  
Integration  should not require substantial development efforts. Similar to service life cycle calls in PaaS clouds~\cite{cloudfoundry}, we must define simple, yet generic interfaces that can be easily implemented by service providers.
In particular,
we prototyped the QoX interpreter aspect illustrated in Figure~\ref{fig:qoxsharing} (outbound logic), though
the actual implementation is a decentralized one,
and in a similar way, integrate into existing systems to use the rating and perform certain actions (inbound logic), including the special case of
monitoring one's own rating. 

The Quality Interpreter is implemented as a set of light weight process that interact with interfaces of other components. It can be configured with a text configuration file that specifies the following: (i) a list of components, each of which has the name of the component, IP, type (service, executor,
sensor), component description and its tasks; (ii) a mapping table that maps monitors feedback to list of actions. 


\subsection{Outbound logic}
The outbound logic extracts and collects information from alerts, events, and status updates from the sensors and sends this information to the Quality Interpreter
for translation into a rating. 


\nip{Intrusion Detection System:} An intrusion detection system (IDS) monitors network traffic and looks for signatures within packets or performs behavioral analysis of the traffic to detect anomalies.  In this case, the shim's outbound logic is designed to intercept the alerts from the IDS, and allow service providers to configure the feedback weights for each alert type. We integrated a shim interface to Snort~\cite{snort}. Snort alerts are configured to log to Syslog.  By using SWATCH~\cite{swatch} to monitor Syslog, the shim is alerted to all Snort alerts. The shim parses the alerts and extracts information such as source IP and alert type and send the feedback to the Quality Interpreter.

\subsection{Inbound logic} 
The shim's inbound logic interprets how incoming ratings should impact the execution of the component. 
%
%
%
%

\nip{Load Balancer:}
In Web services, load balancers are used to distribute
client load across identical instances of the service.  Typically,
load balancers aim for even distribution~\cite{nginx}. With rating information obtained through the informaiton exchange, the web service provider can differentiate its users based on their (expected) QoC, directing good/trusted clients to a set of
servers, and bad/untrusted clients to a different set of servers. We integrated a shim interface with HAProxy load balancer. This shim alters the configurations written in a haproxy.cfg file to specify load balancing based on the rating of consumer.  Upon every change, the shim will tell HAProxy to reload the configuration.

%

\nip{Paas Broker}
PaaS clouds offer the ability
to compose applications using services provided by the platform.  In
some cases, a PaaS cloud provides all of the services (\eg
Microsoft Azure~\cite{windowsazure}).  In other cases, platforms, such as
CloudFoundry~\cite{cloudfoundry}, provide an environment where many
service providers can offer their services.

Service consumers use a {\em broker} to (1) discover services and (2) bind to them. Service
discovery, in general, implements a simple search capability, focusing
on returning one-to-one match with the needed service (\eg version 2.6
of MongoDB).  We extended the CloudFoundry client-side broker interface to enrich and filter the results. Whenever a search request arrives at the broker, the integrated interface would interpose on the request and queries the information exchange service to get rating of the searched services; the inbound logic would then sort and filter the results based on user-defined criteria -- \ie configured to filter out any services that would have low rating, and sort the remaining results based on rating.

\subsection{Ratings Monitor}

\nip{Sentiment Monitor:}
Infrastructure monitoring tools traditionally present information about the infrastructure to the administrators -- while this could be used for outbound logic, we look at this as an interface to monitor the ratings in the information exchange system. Monitoring can be used as a way to alert administrators of changes in their service's rating. This would be implemented in the shim's inbound logic. We took advantage of JNRPE (Java Nagios Remote Plugin Executor)~\cite{jnrpe} to build a Java plugin that is listening for ratings changes, and displays this sentiment and configures alerts for when sentiment (collective metrics of the service provider running Nagios) drop.

\section{Proof-of-concept Evaluation}

\begin{figure}
\centering
\includegraphics[width=\columnwidth]{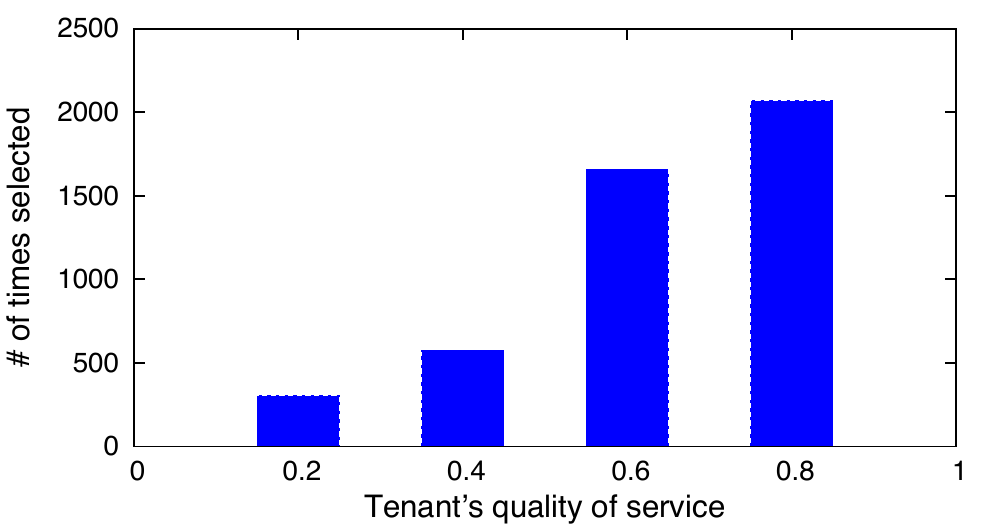}
\caption{Revenue increase for high quality service providers}
\label{fig:revenue}
\vspace{-0.1in}
\end{figure}

As an example to demonstrate the tangible benefits for sharing information, we focus on sharing information about service providers
and how this incentivizes services providers to provide good service. 
We emulated a typical cloud environment, using Mininet~\cite{mininet}, along with the CloudFoundry integration.  For this, 
The service consumer will choose among services that have the highest quality of service. To distinguish between service providers, we use four discrete rating
values -- 0.2, 0.4, 0.6, and 0.8 (higher is better). The service consumer will choose among the top results with some probability distribution (1st search result chosen 85\% of the time, 2nd result 10\% of the time, 3rd result 5\% of the time).
As shown in Figure~\ref{fig:revenue}, the expected benefits hold. As tenants with
the greatest tenant quality (0.8) had a greater revenue, while tenants with lowest tenant quality had the least revenue.

%% file: conclusion.tex
\section{Conclusion}
We presented initial thoughts on including quality of consumption and quality of service in cloud-based services. We discussed the major
challenges that arise when representing and controlling the interactions among service providers and consumers. 
With the proliferation of specialized services and a
growing number of applications, we need to go beyond simply measuring
and reacting, but to share the information with other consumers or
providers. We also discussed how to leverage the (IaaS or PaaS) cloud provider as
a vouching authority to deal with sybils and lying. Further, we showed that the feasibility of integrating QoX in cloud-based services. 

As this is preliminary, there is much to do.  We presented an overarching vision of measuring and sharing QoX information, but were
only able to prove out a subset.
As future work, we plan to investigate and research a number of challenges, such as (i) a general specification language for the interpreting QoX information such that it can feed into the rating system and be used to guide automatic use of ratings to configure executors (our initial interpreter is very basic),  (ii) extend the interpreter by using processing systems to extract structure from unstructured text reviews such that we can automatically incorporate the finer grained details of those reviews, (iii) perform a more in-depth study of service and consumer features to enable automated personalization of results, and (iv) explore the additional information a cloud provider can measure such that it can be used to verify a record of event, without specifically looking for that event.